\documentclass[lnbip]{svmultln}
\usepackage{amsmath}
\usepackage{amsfonts}
\usepackage{amssymb}
\usepackage{graphicx}
\usepackage{epstopdf}
\usepackage{amsmath}
\usepackage{amssymb}
\usepackage{caption}
\usepackage{float}
\usepackage{comment}
\usepackage{graphicx}
\usepackage{hyperref}

\begin{document}
\mainmatter              % start of the contribution
\title{Reprogramming Matter, Life, and Purpose}
\titlerunning{Reprogramming Matter, Life, and Purpose} % abbreviated title (for running head)
\author{Hector Zenil\inst{1,2,3}
%\and
%Elsa Bertino
}
\authorrunning{Hector Zenil}   % abbreviated author list (for running head)
\tocauthor{Hector Zenil}
\institute{Algorithmic Dynamics Lab, Unit of Computational Medicine, Center for Molecular Medicine, SciLifeLab, Karolinska Institute, Stockholm, Sweden.
\and
Department of Computer Science, University of Oxford, Oxford, U.K.
\and
Algorithmic Nature Group, LABORES, Paris, France.
\email{hector.zenil@algorithmicnaturelab.org},\\ Homepage:
\href{http://hectorzenil.net}{http://hectorzenil.net}}

\maketitle              % typeset the title 

\begin{abstract} Reprogramming matter may sound far-fetched, but we have been doing it with increasing power and staggering efficiency for at least 60 years, and for centuries we have been paving the way toward the ultimate reprogrammed fate of the universe, the vessel of all programs. How will we be doing it in 60 years' time and how will it impact life and the purpose both of machines and of humans?

\end{abstract}

\section{Information and Purpose}

%Predicting the future, particularly the future of technology and computation, immediate or over the long term, has typically been a risky proposition. In 2004, Bill Gates predicted that the spam problem would be solved in 2 years (current estimations of spam go from 50\% to up to 90\% of total emails exchanged in the world). Former IBM president, Thomas Watson, is allegedly attributed a prediction from 1943 that there would be a market for only about five computers in the future. This is not to mention  misguided predictions by those running the financial world, such as the CEO of Fannie Mae, Franklin Raines, who insisted that its subprime assets were riskless or the supposed regulator, the Federal Reserve of the U.S., failing to forecast the recession of 2008.

Instead of attempting guesses, let alone answers, about the future, I think the best approach to the future is to ask questions. What will be the purpose of human beings? Of machines? What will be the fate of people rendered redundant when their jobs are automated? These questions driven by business and productivity will determine how we shape our technological future, indeed how we shape our future as such. 

\begin{figure}[htpb!]
\centering  
\includegraphics[width=6cm]{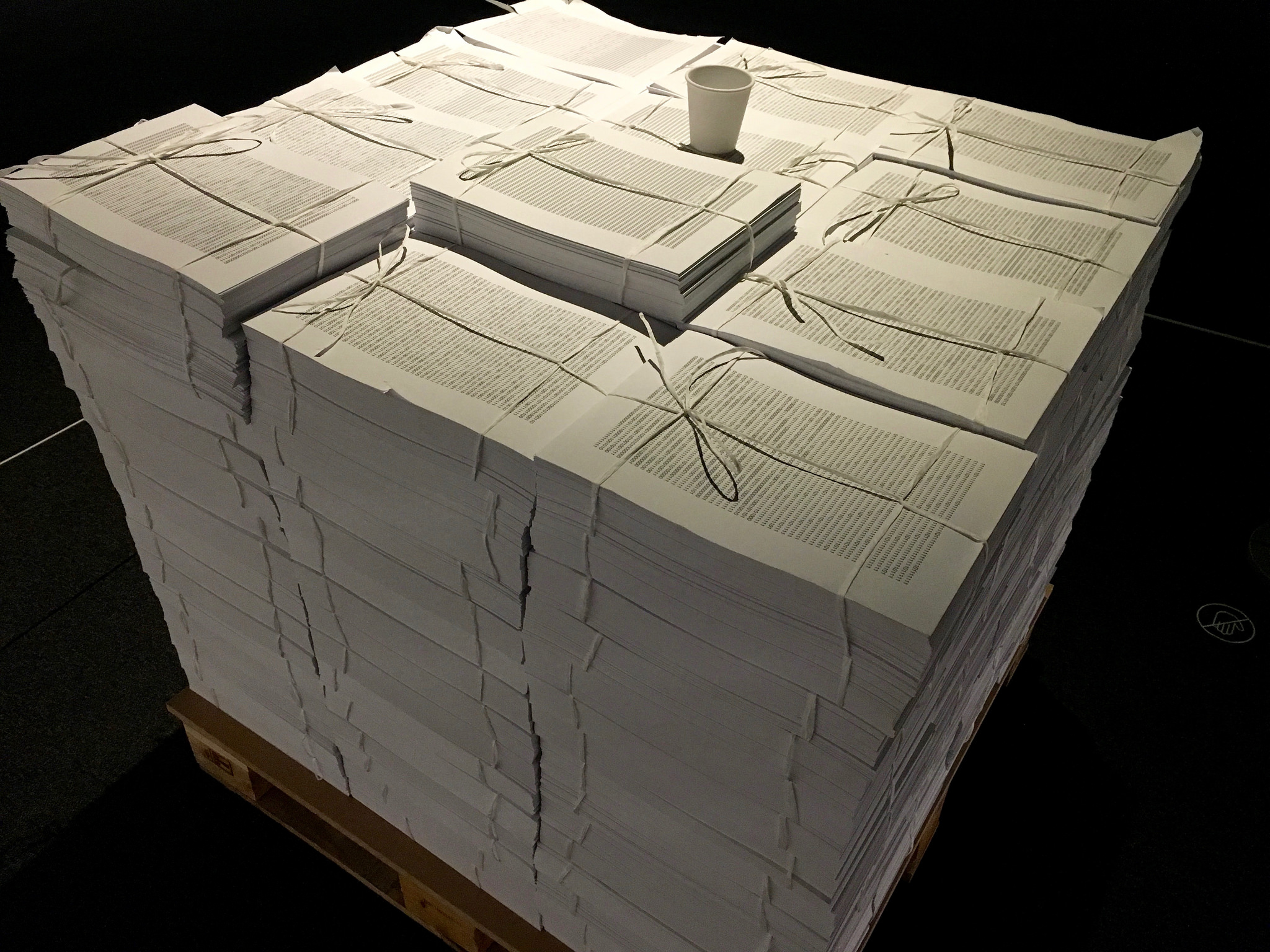}\hspace{.5cm} \includegraphics[width=3.8cm]{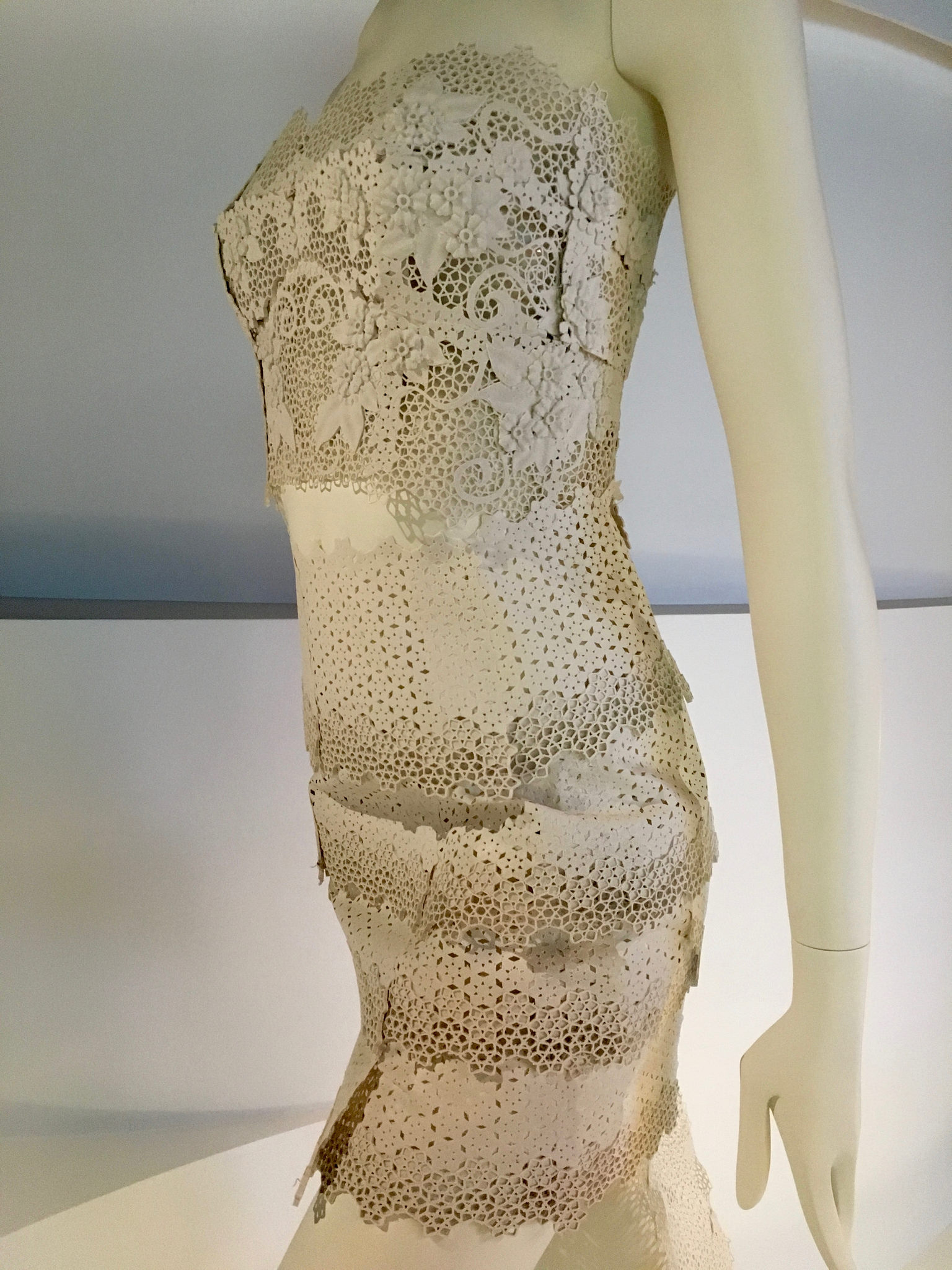}\\ \vspace{.5cm}
\includegraphics[width=6cm]{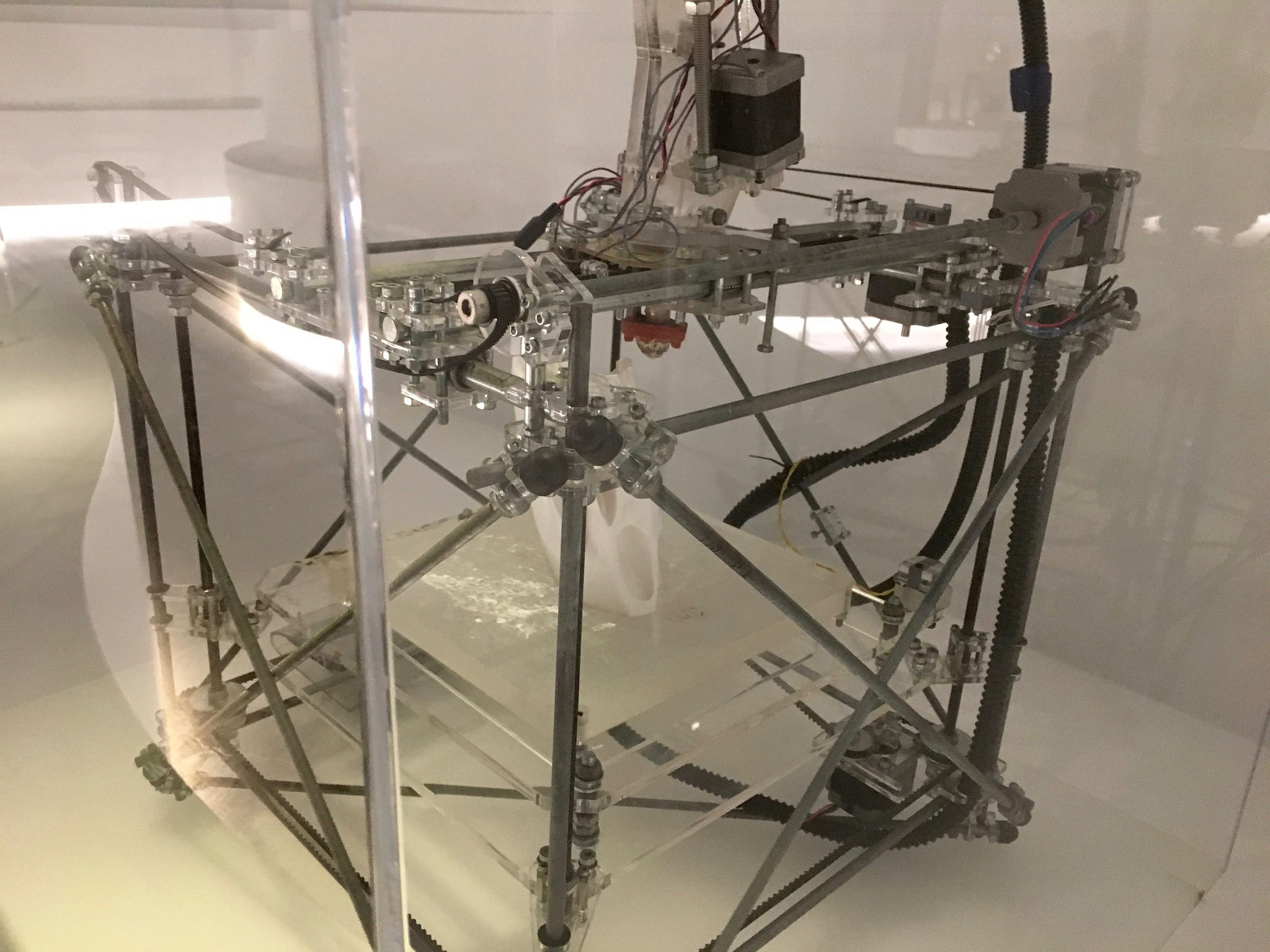}
\caption{
\label{cup}\textbf{Top left:} A pile of paper containing the printed binary computer code with the instructions and computer program for making the 3D-printed cup on top. The length of such a code is a measure of its information content, the content that determines its shape and thus its purpose. The less compressible an object, the more random and less meaningful. Highly structured crafted objects have a high information content; their 3D printed code is large and requires a lot of calculation to produce. In a world of unlimited choices, randomness is tantamount to oblivion while computation frames purpose. \textbf{Top right:} In Bloom 3D printed dress (2014, XYZ Workshop, Melbourne) made of flexible biodegradable plastic. If you want to send a dress to a remote place as a gift you can just send the 3D printing computer program of the dress and it will be reconstructed at the other end. Once printed, the only value of the first dress is to have been printed first. As digital copies do not lose any information and are exact copies of the original, nor is there any human craft involved beyond the design, the boundary of original and copy blurs. \textbf{Bottom:} A 3D printer capable of printing its own parts by following its own blueprint to produce another 3D printer, self-replicating machines, a milestone of biological life, towards matter reprogramming matter with no biological intervention, yet purpose is external to the printer. \textit{All pictures taken by HZ at the exhibition 'Out of Hand: Materializing the Digital', Powerhouse Museum---Museum Of Applied Arts And Sciences, Sydney, Australia, 2017.}}
\end{figure}

It is becoming more and more apparent that our attention is not only limited in range but also in depth and time-span. Bombarded with information and misinformation, what to pay attention to has become a central concern, here the concept of mathematical randomness will be key to understand the difference between meaning, value, and purpose (Fig.~\ref{cup}). Machines do not have this problem, they do not have to choose what they pay attention to except when fulfilling human purposes. Resources permitting, machines can afford to pay attention to everything, avoiding all mistakes, innocent of purpose. One of the main challenges is, however, to inject purpose into machines, to make them look focused, attentive or `intelligently' stupid to engage with us. Formal approaches to information will play a key role in framing all sorts of meaningful questions (Fig.~\ref{cup}).

\begin{figure}[htpb!]
\centering  
\includegraphics[width=7.5cm]{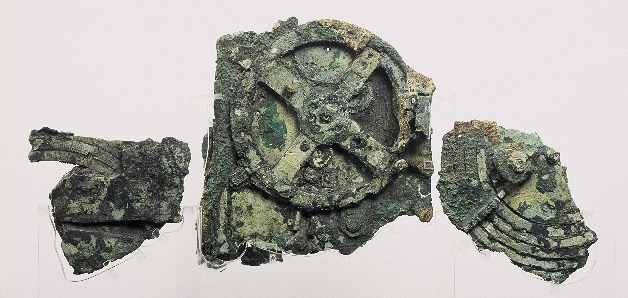}\hspace{.45cm}
\includegraphics[width=4cm]{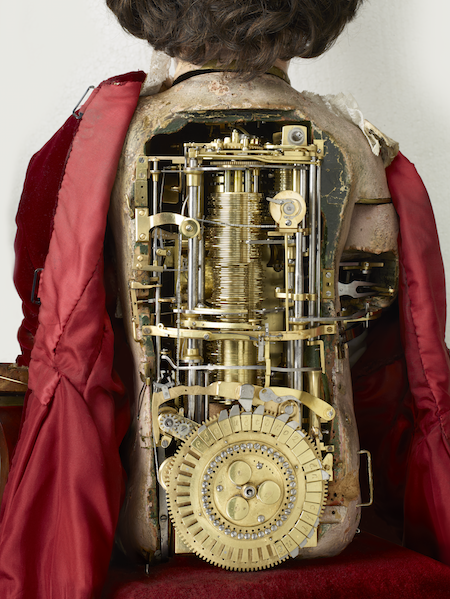}\\
\vspace{.4cm}
\includegraphics[width=5.3cm]{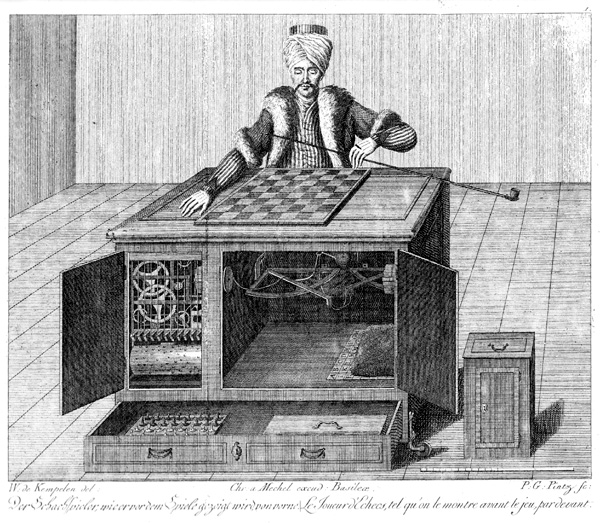}\hspace{1cm}
\includegraphics[width=5.5cm]{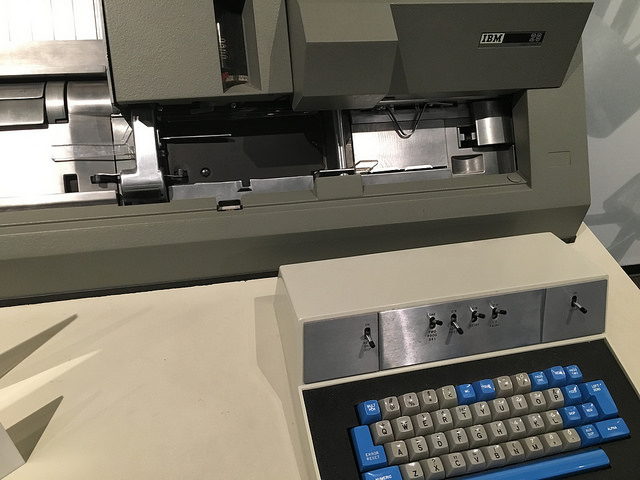}
\caption{
\label{ecrivainturk}\textbf{Top left:} Antikythera Mechanism (inv. no. X 15087: fragments A, B, C. An early astronomical calculator based upon cogs that would be reinvented almost 2 thousand years later in a case of technological convergence. \textit{Permission from the National Archaeological Museum, Athens (Photographer: Kostas Xenikakis) Copyright \copyright Hellenic Ministry of Culture and Sports/Archaeological Receipts Fund}. \textbf{Top right:} The writer automaton and the mechanical Turk are among the first modern attempts to reprogram matter by rearranging and reshaping other matter such as cogs, pulleys and camshafts to perform a specific task. This is one of 3 automata of this type created by the inventor who inspired the movie Hugo. Permission from the \textit{Mus\'ee d'art et d'histoire, Neuch\^{a}tel}. \textbf{Bottom left:} The Mechanical Turk (\textit{picture in the public domain}~\cite{turk}), a human pretending to be a machine that pretends to be a human in anticipation of the convoluted relation between human and machine, the simulation and the simulacrum. \textbf{Bottom right:} IBM 029 Card Punch machine capturing instructions provided by humans for machines to follow. \textit{(By HZ at the Living Computers Museum, Seattle, WA)}.
}
\end{figure}

One of the subjects I study, algorithmic probability, suggests that things are much less disconnected than classical probability usually presupposes, that events are not independent of each other and that what can happen will happen more often than predicted (e.g. market crashes). So when I consider our technological future I cannot decide whether things will go worse for humanity because they can, or whether things will go better than expected because our sense of what can go wrong is exaggerated. We can play it safe by looking at what is already happening, even though we may not fully grasp the present moment in all its multifacetedness. For example, people tend to think that machines will replace human jobs, but they have been doing so for the last 50 years or so. People also fantasize about the future of our virtual human lives, when we already favour online social interaction over physical. People know that humans started creating and using tools to advance and flourish in a human-dominated world, but wonder what humanity will look like when dominated by computers. However, it is difficult, if not impossible, to conceive most current human activity without computers. My focus here will only be on things that are already occurring, and on how these developments fit into the story of humanity's creation of tools for shaping its world. My interest is in how humans team up with computers to remake the world and how humans have reprogrammed the world starting as early as we started using tools and putting sophisticated moving parts together (see top left of Fig~\ref{ecrivainturk}).

Computers are small, human-controlled, reprogrammed space-time regions, pieces of the universe which we have diverted from their usual course and put to work performing computations that serve our particular ends. They are an example of the living reprogramming the inanimate (see bottom left of Fig.~\ref{ecrivainturk}), and we are now also witnessing the living reprogramming the living, generating unimagined power (cell reprogramming). This power will bring in its wake a world of opportunities (and new responsibilities) only comparable to our wildest fantasies and beyond any known sci-fi movie scenario, a world that we cannot conceive of or understand from today's vantage point. 

\begin{figure}[htpb!]
\centering  
\includegraphics[width=5cm]{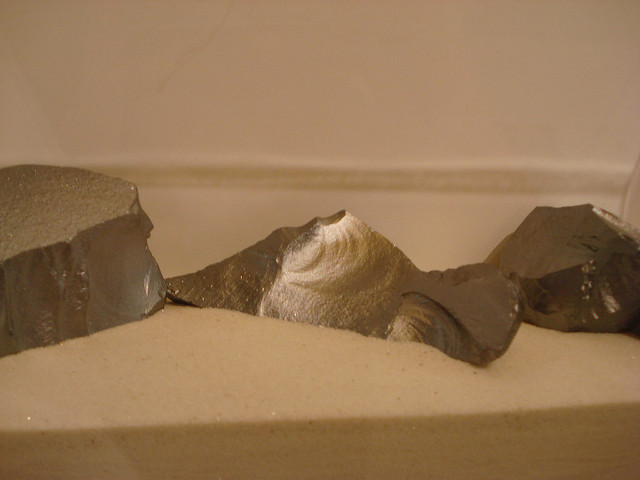}\hspace{1.6cm} \includegraphics[width=3.4cm]{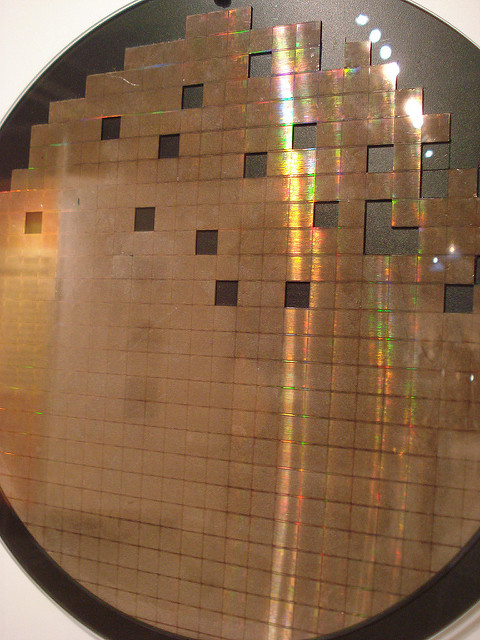}\\
\vspace{.5cm}
\hspace{.7cm}\includegraphics[width=3.6cm]{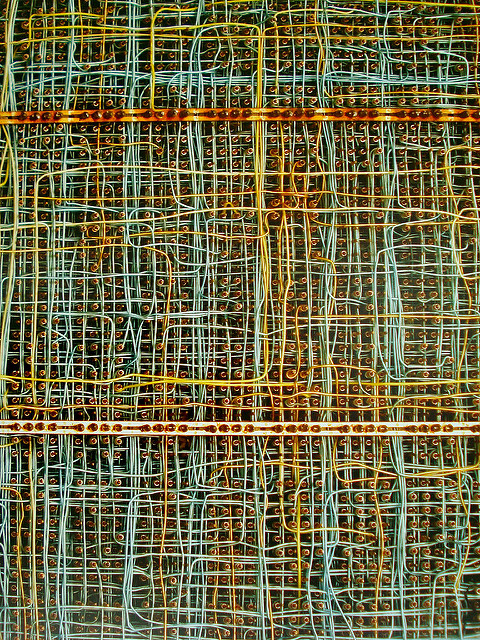}\hspace{1.6cm}\includegraphics[width=5cm]{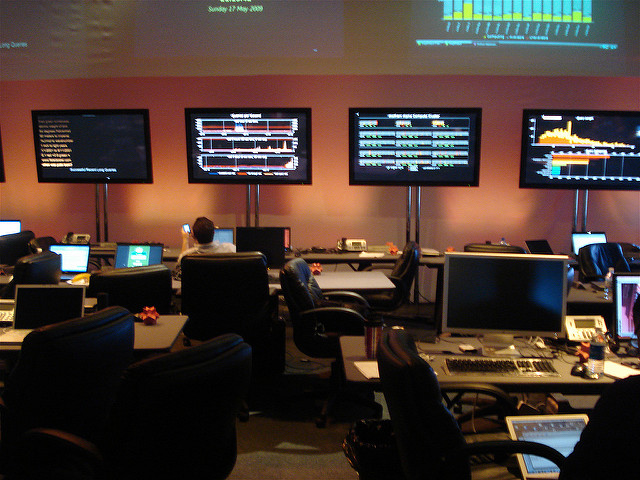}
\caption{
\label{rawmaterials}Rewiring the world from a naturally occurring element (raw silicon) to performing the most sophisticated calculations. Controlling energy at the lowest scales in the way electrons and photons travel in cable lines. \textbf{Top left:} Raw silicon as extracted from Silicon Valley displayed at the Intel Museum, CA \textit{(By HZ)} and transformed into the `naturally binarizing' chemical element which is just sensitive but also robust enough to convert continuous electrical signals into stable binary choices, letting current pass or not depending on a sharp threshold. \textbf{Top right:} Silicon wafers at the Intel Museum contain enormous numbers of microprocessors ready to do our calculations, borrowing matter and energy from the universe. \textit{(By HZ)}. \textbf{Left bottom:} Another example of rewiring the world: a server rack at the Intel Museum \textit{(By HZ)}. \textbf{Right bottom:} Humans monitoring in real time the result of reprogramming the world on the launch day of \textit{WolframAlpha}, the most sophisticated computational knowledge engine that produces original answers to natural language queries \textit{(Picture in the control room, taken by HZ, University of Illinois Urbana-Champaign)}.
}
\end{figure}

Over the course of its history, humanity has had increasing success mastering the flow of energy---early in human history by producing and controlling fire, and today by controlling the flow of electrons and photons in modern electronic computers. From raw materials provided by our planet (see Fig.~\ref{rawmaterials}) we have extracted carbon, silicon and metals and used them to build artifacts that borrow energy from the sun and effectively reprogram small regions of the universe inside boxes called electronic computers, artifacts made of atoms and electrons which would otherwise be elsewhere, performing other functions and following the normal course of the universe.

\section{Shape Is What Tells Us Apart}

Reshaping matter at the lowest scale is achieved by rearranging and rewiring atoms into naturally unlikely but mathematically stable structures with unprecedented properties such as strength, flexibility and thickness, properties that are practically impossible to find in nature and that allow new applications of nanomaterials in our everyday lives. When absorbed by cancer cells and exposed to light radiation, one of these designed elements (Buckminsterfullerene C60) damages the DNA, proteins, and lipids of the cancer cell, forcing it to go into reprogrammed cell-death (called \textit{apoptosis}). 

Medical research will benefit greatly from the power of molecular reprogramming. For example, in 2006 a group of Japanese scientists found a way to activate 4 genes from skin cells and convert them into embryonic stem cells (ESCs). These researchers were awarded the Nobel Prize for Physiology or Medicine by the Karolinska Institute. ESCs are cells that can become almost any other possible cell of the nearly 200 different tissue cells that go into building a human being, serving as the basic building blocks out of which are fashioned cells of the heart, brain, muscle and liver, among others.

One problem we face today is that while we know how to reprogram cells, we are hardly able to determine their fate e.g. to determine with some precision what type of cell a stem cell will become. When used in regenerative medicine to fight diseases or help the body heal, most of the time they become dangerous cancerogenous cells instead. This has been the ultimate proof that molecular biology is more similar to digital computation than anyone could have expected. Not only is the code written in a 4-digit discrete programming language, but the code actually implements a computer program that the cell blindly follows. Genes are like subroutines, one can switch them from one place to another and even exchange them across species in an extraordinary modular fashion. The landmark experiment in the 70s that replaced a leg with a hand in a salamander by exchanging a gene marked the beginning of this development. 

Naturally differentiating cells use almost conventional communication channels to coordinate actions and build up tissues. Cytokines are the chemical messages on which natural reprogramming occurs, and one of the many handles that can be harnessed to artificially reprogram cells, as nature currently does.  DNA is like a computer program, the cell being the machinery that follows the instructions with the help of biochemical reactions (e.g. what in biology we call enzymes). At the Karolinska Institute, for example, it has been discovered that placenta cells can replace liver cells and help regenerate heavily damaged livers. Here, reprogramming cells gives them a different purpose. 

More recently, in 2016, Craig Venter built a 500-gene organism written from scratch in his quest to find the shortest code for life, thereby blurring the line between the inanimate and the animate, matter and life--synthetic life. The smallest synthetic organism was the first to be isolated from the `tree' (or web) of life that connects us all to the very first forms of life on Earth. Rearranging and reshaping life at the lowest level is occurring as we speak, with the FDA approving epigenetic drugs. The same lab that produced the first synthetic organism also succeeded in transplanting a genome from one species to another, and `booted it up' to convert the host species into another one, effectively reprogramming it. You would have a very different purpose in life if converted into a fish. Reprogramming thus changes everything.

With the use of gene-editing techniques such as CRISPR/Cas9 to modify genomic DNA at will to perform in vivo perturbation analysis at significantly lower costs, we will reach a much better understanding of true causes than we can using regression analysis, an understanding arrived at by using a more model-oriented approach based upon the generation of perturbation data. Finding causes will allow us to reconstruct first principles and generating mechanisms, effectively the computer program for which a set of initial configurations as input will lead to identifiable and convergent outputs, such as a stable attractor in an energy landscape. This will provide us with the ultimate blueprint of life, which may be used to effectively reprogram the basic units that comprise multicellular organisms. In biology, shape is function. Proteins are like Lego blocks that together build up tissues which in turn build up every inch of our bodies. Manipulating the computer programs that produce different shapes by fine-tuning gene expression to produce (different) proteins, we will have the wherewithal to steer life at will.

The first clue that evolutionary biology was somehow algorithmic comes from artificial selection: we have been genetically modifying organisms for about 12000 years, exploiting plants and animals for our own benefit. Then there are the genetic experiments of Gregor Mendel early in the 19th century, which showed not only that information was transferred across generations but that this transmission occurred with near-mathematical precision. A third and even more dramatic milestone was the discovery of the basic structure of life's digital alphabet, DNA, by Crick, Watson, Wilkins and Franklin. It became evident from the modular nature of the genomes of all living organisms on earth that biology was not only highly algorithmic but that nature reprogrammed life like a software engineer. This insight inspired synthetic biologists to switch somatic nuclei across species, producing spider silk from goat's milk and effectively creating a spider-goat, or making plants glow by transplanting genes from luminescent fish.

Thus while competition has been reprogramming life by natural selection, advancements at the end of the last century allowed reprogramming at the cellular level, by hacking a cell with signals that prompted a cell's genetic code to run a different routine than it otherwise would, unless the same signals obtained in the cell's natural environment (e.g. other cells' signals). 

If pioneers such as Alan Turing, Alonzo Church and Emile Post can be considered the first computer programmers, it is Ian Wilmut, Keith Campbell, John Gurdon and Shinya Yamanaka and their colleagues that must be considered the first biological programmers. They not only showed that nuclei could be replaced inside a cell in order to hack the cellular machinery and make it follow the genetic instructions of another organism, but they also identified what are today known as the Yamanaka factors (Oct4, Sox2, Klf4, and c-Myc), a discovery which earned a Nobel Prize. These are signals that act as initial conditions to reprogram adult cells into what are called \textit{induced pluripotent stem cells}, which  have the potential to transform into other cells, thereby becoming any possible tissue of a plant or animal body. Nature does this all the time--during embryonic development signals regenerate entire organs--but by identifying these signals and using them to activate other cells' functions, we make cells behave according to programs put in place by nature but not currently running.

We can now harness biology beyond artificial selection and signal reprogramming, intervening at a lower molecular and nucleotide level, hacking the machinery of molecular biology and its DNA-repair mechanisms to reprogram cells with techniques such as CRISPR/Cas9, which thus becomes a de facto biological reprogramming script that hacks a DNA-repair mechanism to insert a valid instruction in the form of a gene into an otherwise natural occurring DNA sequence.

The protocol that identified the minimally required core set of genes (the Yamanaka factors) that when overexpressed induced pluripotency (in mice and humans) was found by elimination, based upon previous knowledge and trial and error. The process is actually quite inefficient---- both the discovery of these sets and their deployment in reprogramming a cell. So to continue making progress, we need not only the technology to perform genetic interventions but a better and deeper understanding of the causal mechanisms activating a cell's behaviour, revealing its generating mechanisms (their original `computing routines'). To this end, molecular biologists have moved from sequences to networks (see Fig.~\ref{universality}), realizing that sequencing a genome is different to understanding how such a genome actually works. The functional content of the genome is not in the genes themselves, but in the way the genes interact with each other by producing proteins that in turn interact with other genes (or with the originating gene). The Yamanaka factors were later actually found to be connected among themselves and with other genes in a sophisticated fashion, revealing previously unknown signalling and metabolic links. These key interactions are represented by links among functional units (genes, proteins, or signals, among other possible functional units) and my colleagues and I at my labs are developing methods and tools with which to tackle this challenge of discovering causality, laying bare a cell's own software and limning the way it can be reprogrammed.

\section{Reprogrammable Nature}

Way back in 1936, Alan Turing gave us the tools to understand the depth and breadth of the concept that we identify today as `computational universality', that is, the ability of certain computer programs and actual mechanical and electronic computers to perform any task that can be algorithmically described. Your laptop or tablet can play music, plot a graph, or serve as a typewriter. This is because it is reprogrammable; it can emulate any computer program. We know that for a computer to be reprogrammable, very few elements are needed. For example, a tape and a writing head can make a reprogrammable computer, as can DNA with a splicing operation propelled by chemical reactions, achieving computational universality.

But while scientists have known for a while that most computers are universal, until now we have not known to what extent computer programs are similarly universal. By using a technique of emulation~\cite{zenilriedel}, we were able to show that almost every computer program can be reprogrammed. Thus, we demonstrated how pervasive and ubiquitous universality and reprogrammability are~\cite{riedelzenil}. 

We now know not only that the world seems to be fully reprogrammable in an abstract way, but also that the initial conditions of a computer program---such as a deterministic rule or a physical law---are just as important as the program itself. This is because its initial conditions can fundamentally change the typical behaviour of a system, and make it appear arbitrarily more or less complicated~\cite{riedelzenil}. 

What has prevailed today, however, is the construction of dumb devices that minimize intelligence but maximize specific-purpose automation. For example, a dishwasher is a silly box that can hardly be considered sensitive at all, even to the environment that is created in its interior, let alone to an artificial one. But it makes no sense to have general-purpose dish washers. This is changing, a development that is known in the industry as the Internet of Things (IoT), the embedding of interconnected devices capable of sophisticated computation (IoT was actually known in the past as `pervasive computing').

But a universal computer can easily reprogram the dishwasher for some other purpose (subject no doubt to the limitations of its hardware), so there is nothing fundamental about general-purpose intelligence once one reaches the power of universal computation, and deep learning is just evidence of this. Deep learning is flourishing today thanks not to any fundamental breakthrough but because of the availability of data for training purposes and because of the speed and low cost of intensive computation. 

Artificial intelligence does not have to be expensive. The same research that we have undertaken indicates that even apparently silly devices have the potential to be reprogrammed to behave as smart devices. Once computing becomes more ubiquitous in all sorts of devices, these will be able to take the power of the dishwashing machine and transform it, if there is anything interesting that can be done with it. A better example is the fridge, yet another dumb box at home, a fixture of the kitchen. It does not take much to make a fridge smarter, just as it takes almost nothing, apart than a small device such as Apple TV or Amazon Fire, to transform a silly screen into a smart TV. This is the power of universal computing and of reprogramming.

In the year 2065 and beyond, as we already do today with laptops, tablets, smartphones and smartwatches, we will run apps on a wider range of devices, and these devices will increase further the interconnectivity among things, making everything considerably smarter. Almost everything we use will have a powerful electronic CPU capable of universal computation and therefore of reprogramming the device in which it resides. Because everything will be a reprogrammable computer, we will be even more capable and effective at reprogramming matter and shaping the universe that surrounds us, making us ask again who we are and what we will become. 

\section{Ubiquitousness of Universal Computation}

Classical mechanics describes the world, and determines it---in terms of computer programs derived from fully deterministic laws, and of inputs in the form of specific physical initial conditions. This means that we can understand, but also try to manipulate these laws and the universe itself in terms of computation. In biology, for example, this realization will have far-reaching consequences. In a simplification of things, cells, for example, can be represented by small programs in the form of networks of genes that turn on and off other genes, which in turn activate or reduce the production of living systems' building blocks, proteins. By manipulating genes and proteins one can steer the program and make the cell behave differently, or even become something different. 

By introducing methods based upon the theory of computation and information, my team and I manipulate biological programs such as genetic networks, evaluating and then moving the networks toward or away from randomness, to, for example, eventually destabilize a cancer cell and make it ineffective~\cite{zenilschmidttegner}. In the future, it seems clear that we will be able to repair or dispose of cells before we even develop symptoms, and we will do so directly at the source and at the right scale, cellular or even atomic, a scenario that brings to mind Star Trek's medical `tricorder', the portable medical device used to diagnose and cure in the sci-fi series. Today we have literally taken the calculating power of a small region of the universe, living and inanimate, and used it to perform our very own calculations. Imagine what we will be able to do in the future if we manage to become more efficient at exploring and harnessing the sources of the Sun's and the Earth's energy. This minor deviation from the otherwise regular computational course of the universe may look small at the cosmic scale, but it is a huge accomplishment that may end up reshaping the universe itself, and even creating other universes.

\begin{figure}[htpb!]
\centering
\includegraphics[width=5.5cm]{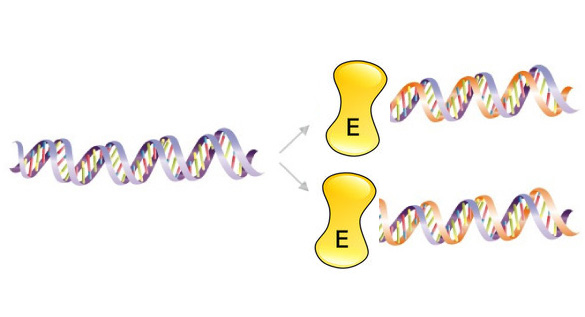}\hspace{1cm}\includegraphics[width=5.5cm]{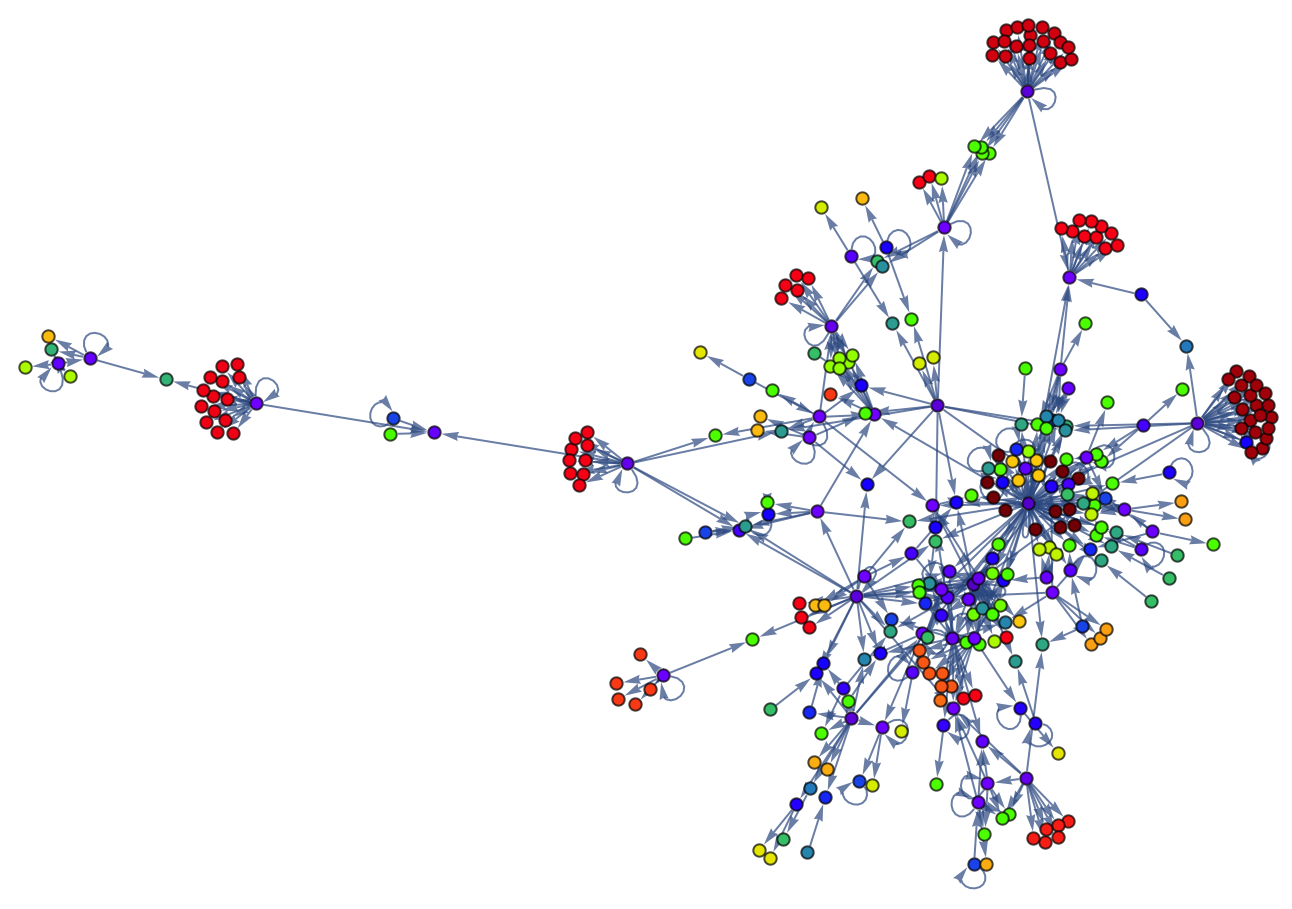}\\
\vspace{.6cm}
\includegraphics[width=5.8cm]{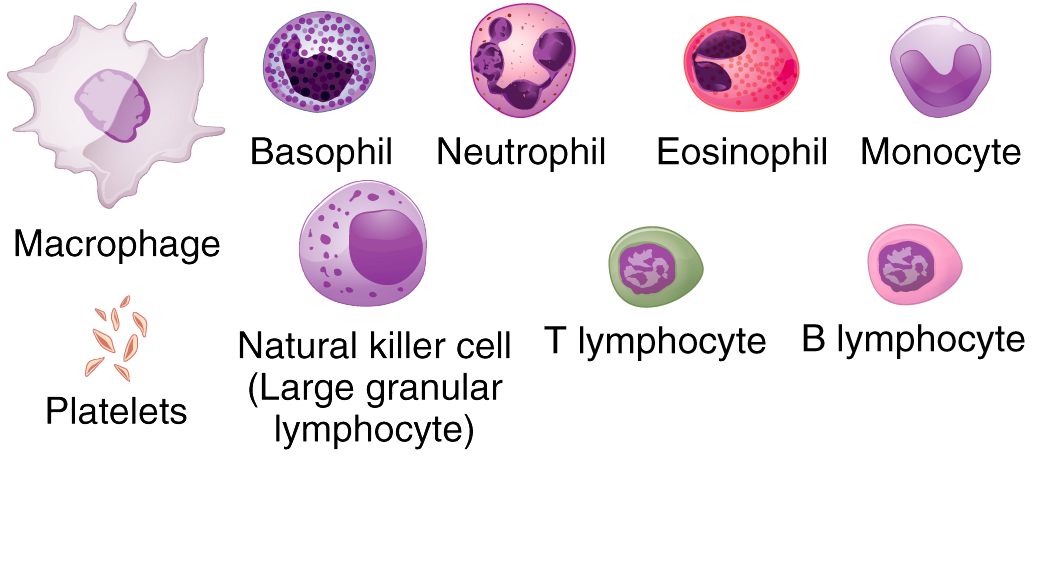}\hspace{.7cm}\includegraphics[width=5.4cm]{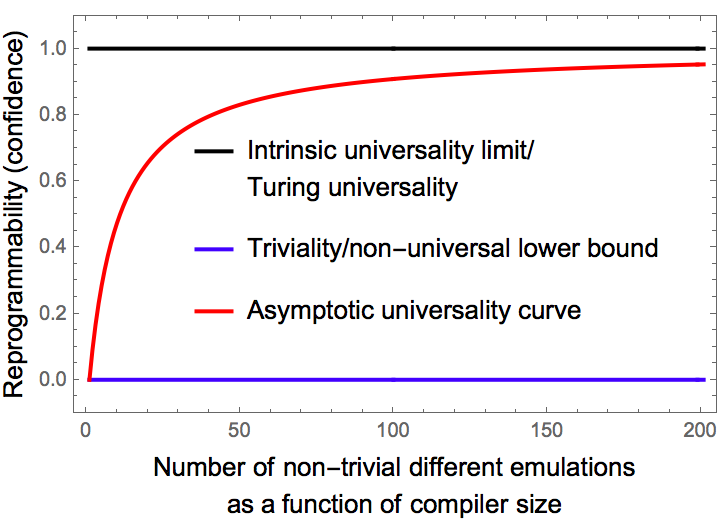}

\caption{
\label{universality}\textbf{Top left:} Error correcting mechanisms en route  from copying to transcription and translation. A chemical machine called an \textit{enzyme} performs a basic operation similar to a finite automaton verifying a regular expression. While individual enzymes may not be computationally very sophisticated, there are dozens of enzymes for different operations that together constitute a fully computational device. \textbf{Top right:} Functional operation between genes from indirect physical interaction (proteins, but also various forms of RNA messages) are represented by links connected in non-random fashion. Depicted here for purposes of illustration is the Transcription Factor (TF) network of e.coli, one of the most studied living organisms in biology. TF is the name given to a gene that can regulate other genes. \textbf{Bottom left:} The way in which the genes interact gives rise to different cells, of which there are about 200 basic types (or tissues) in the human body, each devoted to a particular function that contributes to the harmony of a multicellular living organism. Depicted here is a subset of important blood cells that interact with each other and belong to the immune system, a debugging system that literally gets rid of `bugs' and `buggy cells'. Rewiring the underlying genetic network transforms one cell into another. \textbf{Bottom right:} Number of abstract programs that can be reprogrammed, increasingly simulating a greater number of other computer programs that behave very differently, qualitatively speaking, thereby indicating the pervasiveness of computation and the ubiquity of Turing-universality. We introduced a concept of statistical evidence to quantify the chances of a computer program being Turing-universal, increasing them every time that such a computer program emulated another qualitatively different computer program, thus effectively providing a numerical measure of reprogrammability. When one considers all possible computer programs sorted from smaller to greater length in bits, coupled with increasing lengths of initial conditions that act as translators (or compilers), it becomes clear that all computer programs are Turing-universal with asymptotic probability 1 (red line).}
\end{figure}

My colleagues and I have proven that almost anything that looks reprogrammable is reprogrammable (see Fig.~\ref{universality}). At the smallest scale, this means that anything can simulate anything else. In the universe, everything is made of the same particles, only arranged in a different way and following a different dynamical, reprogrammable, path. This is not only true for some processes but, in principle, for any process that is the result of rule-based computation, so the implications of the results in the real world may be far-reaching.

An open question is the nature of intelligence. If the mind is the result of the computation of the brain, there is nothing that prevents the creation of a mind, and thus everything that may derive from it, from general-purpose intelligence to consciousness itself. What is obvious today is that we have teamed up with computers to reprogram matter and life, and we cannot easily explain ourselves without computers and what computers do for us, which is much better than what we can do for ourselves in terms of calculations and tasks.

By outsourcing doing and thinking outside the human brain and mind, we will turn inanimate matter into an extension of ourselves, increasingly blurring the boundaries between the two. Further into the future, intelligence will be embedded all around and beyond us in a world without clear boundaries between who or what is thinking and who has directed them to think. We have already outsourced our collective memory (e.g. in books, and today to a significant extent in digital form) but also our personal memory (in digital pictures, blogs, email and so on). We keep outsourcing cognitive functions such as navigation. In the future we will continue delegating cognitive functions to the point that the line between us and our computing assistance in the cloud will be difficult to distinguish, other than by the consciousness that characterizes us, assuming this cannot ever be outsourced. It seems that a teleological purpose of intelligent beings is to reprogram their surrounding world, but until now we have not known the extent to which this would be possible, both in terms of human willingness and the amenability of matter and energy to being manipulated to perform computation. 

We are just starting to scratch the surface of the many possibilities that universal computation has opened up for the future in the way of reprogramming matter and life. And our current efforts to reprogram it actively contribute to shaping the final computational fate of the universe itself.


%
\end{document}